\documentclass[preprint,aps,showpacs,nofootinbib]{revtex4}

\usepackage{epsfig,amssymb,amsmath}

\def\bea{\begin{eqnarray}}
\def\eea{\end{eqnarray}}
\def\beq{\begin{equation}}
\def\eeq{\end{equation}}

\newcommand{\bear}{\begin{array}}
\newcommand {\eear}{\end{array}}
\newcommand{\bef}{\begin{figure}}
\newcommand {\eef}{\end{figure}}
\newcommand{\bec}{\begin{center}}
\newcommand {\eec}{\end{center}}

\def\GEV#1{10^{#1}{\rm\,GeV}}

\begin{document}
\draft
\tighten
\preprint{TU-899}
\preprint{IPMU12-0010}
\title{\large \bf
Light Higgsino from Axion Dark Radiation}
\author{
    Kwang Sik Jeong$^a$\footnote{email: ksjeong@tuhep.phys.tohoku.ac.jp}
    and
    Fuminobu Takahashi$^{a,b}$\footnote{email: fumi@tuhep.phys.tohoku.ac.jp}}
\affiliation{
    $^a$Department of Physics, Tohoku University, Sendai 980-8578, Japan\\
    $^b$Institute for the Physics and Mathematics of the Universe,
    The University of Tokyo, \\ 5-1-5 Kashiwanoha,
    Kashiwa, Chiba 277-8582, Japan
    }
%\date{\today}

\vspace{2cm}

\begin{abstract}
The recent observations imply that there is an extra relativistic degree
of freedom coined dark radiation.
We argue that the QCD axion is a plausible candidate for the dark radiation,
not only because of its extremely small mass, but also because in the
supersymmetric extension of the Peccei-Quinn mechanism the saxion tends
to dominate the Universe and decays into axions with a sizable branching
fraction.
We show that the Higgsino mixing parameter $\mu$ is bounded from above when
the axions produced at the saxion decays constitute the dark radiation:
$\mu\lesssim 300$ GeV for a saxion lighter than $2m_W$,
and $\mu$ less than the saxion mass otherwise.
Interestingly, the Higgsino can be light enough to be within the reach of LHC
and/or ILC even when the other superparticles are heavy with mass about 1
TeV or higher.
We also estimate the abundance of axino produced by the decays of Higgsino
and saxion.
\end{abstract}

%\pacs{}
\maketitle

%\tableofcontents
%\newpage

\section{Introduction}

The present Universe is dominated by the dark sector, i.e., dark matter
and dark energy, although it is not yet known what they are made of.
Therefore, it may not be so surprising if there is another dark component,
which behaves like radiation.

The presence of additional relativistic particles increases the expansion
rate of the Universe, which affects the cosmic microwave background (CMB)
as well as the big bang nucleosynthesis (BBN) yield of light elements,
especially $^4$He.
The amount of the relativistic particles is expressed in terms of
the effective number of light fermion species, $N_{\rm eff}$, and it is
given by $N_{\rm eff} \approx 3.046$ for the standard model.
Therefore, if $N_{\rm eff} >  3$ is confirmed by observation, it would
immediately call for new physics.

Interestingly, there is accumulating evidence for the existence of additional
relativistic degrees of freedom.
The latest analysis using the CMB data (WMAP7~\cite{Komatsu:2010fb} and
SPT~\cite{Dunkley:2010ge}) has given $N_{\rm eff} = 3.86 \pm 0.42$
(1$\sigma$ C.L.)~\cite{Keisler:2011aw}.
Other recent analysis can be found in Refs.~\cite{Komatsu:2010fb,Dunkley:2010ge,
Hou:2011ec,GonzalezMorales:2011ty,Hamann:2011ge,Archidiacono:2011gq,Hamann:2011hu}.
The $^4$He mass fraction $Y_p$ is sensitive to the expansion rate of the Universe
during the BBN epoch\footnote{
$Y_p$ is also sensitive to large lepton asymmetry, especially of the
electron type, if
any~\cite{Enqvist:1990ad,Foot:1995qk,Shi:1996ic,MarchRussell:1999ig,Kawasaki:2002hq,
Yamaguchi:2002vw,Ichikawa:2004pb}.
},
although it has somewhat checkered history since it is very difficult to estimate
systematic errors for deriving the primordial abundance from $^4$He
observations~\cite{Olive:2004kq}.
Nevertheless, it is interesting that an excess of $Y_p$ at the $2 \sigma$ level,
$Y_p = 0.2565 \pm 0.0010\, ({\rm stat}) \pm 0.0050\, ({\rm syst})$, was reported
in Ref.~\cite{Izotov:2010ca}, which can be understood in terms of the effective
number of neutrinos, $N_{\rm eff} = 3.68^{+0.80}_{-0.70}$ $(2 \sigma)$.\footnote{
The authors of Ref.~\cite{Aver:2010wq} estimated the primordial
Helium abundance with an unrestricted Monte Carlo taking account of
all systematic corrections and obtained $Y_p = 0.2561 \pm 0.0108$\,$(68\%{\rm CL})$,
which is in broad agreement with the WMAP result.
}
Interestingly, it was recently pointed out that the observed deuterium abundance
D/H also favors the presence of extra radiation~\cite{Hamann:2010bk,Nollett:2011aa}:
$N_{\rm eff} = 3.90 \pm 0.44$ $(1 \sigma)$ was derived from the CMB and
D/H data~\cite{Nollett:2011aa}.
It is intriguing that the CMB data as well as the Helium and Deuterium abundance
favor additional relativistic species, $\Delta N_{\rm eff} \sim 1$, while they are
sensitive to the expansion rate of the Universe at vastly different times.

The extra radiation may be dark radiation composed of unknown particles.
Then it is a puzzle why it is relativistic at the recombination epoch, why the
abundance is given by $\Delta N_{\rm eff} \sim 1$, and why it has very weak
interactions with the standard-model particles.

In fact, there is a well-motivated particle with the desired properties, namely,
the QCD axion.
The axion appears in the Peccei-Quinn (PQ) mechanism~\cite{Peccei:1977hh}, one
of the solutions to the strong CP problem~\cite{Kim:2008hd}, in association with
the spontaneous breakdown of the PQ symmetry.
The axion remains extremely light: its mass is originated from the QCD anomaly
and is in the range of $10^{-5} {\rm\, eV} \lesssim m_a \lesssim 10^{-3}$\,eV
for the PQ breaking scale in the cosmological window, $F_a = \GEV{10} \sim \GEV{12}$.
Furthermore, in a supersymmetric (SUSY) framework, the saxion tends to dominate
the energy density of the Universe, and it decays mainly into a pair of axions.
Such non-thermally produced axions naturally remain relativistic until present.
Thus, the axion is a plausible candidate for the dark radiation.\footnote{
The abundance of relativistic axions produced by flaton decays was studied
in detail in Ref.~\cite{Chun:2000jr}.
The late-time increase of $N_{\rm eff}$ by decaying particles (e.g. saxion to
two axions, and gravitino to axion and axino) was studied in
Ref.~\cite{Ichikawa:2007jv}.
The light gravitinos produced by inflaton decay can account for dark radiation
in analogous to Ref.~\cite{Takahashi:2007tz}.
See also recent
Refs.~\cite{Hasenkamp:2011em,Menestrina:2011mz,Kobayashi:2011hp,Hooper:2011aj}.
The possibility that the $X$ particle was in thermal equilibrium was studied in
Ref.~\cite{Nakayama:2010vs}.
}

In this paper we consider the QCD axion as a candidate for the dark radiation,
and show that the branching fraction of the saxion into axions naturally falls in
the right range, if the axion multiplet is coupled to the Higgs superfields
as in the Dine-Fischler-Srednicki-Zhitnitsky (DFSZ) axion
model~\cite{Dine:1981rt,Zhitnitsky:1980tq}.
Most important, we find that  the $\mu$-parameter is bounded from above,
namely $\mu\lesssim 300$ GeV for the saxion lighter than $2m_W$, and $\mu$ less
than the saxion mass otherwise, independent of the PQ scale $F_a$.
As we shall see later, if the radiative corrections play an important role in
the stabilization of the saxion, its mass can be naturally smaller than the soft
masses for the SUSY standard model (SSM) particles.
This implies that, even if the other SUSY particles are so heavy that
they are above the reach of the Large Hadron Collider (LHC), the Higgsino should
remain within the reach of the LHC and/or future collider experiments such
as International Linear Collider (ILC).
This will be of great importance especially if the mass of the SM-like Higgs boson
is indeed around $124 - 126$\,GeV as suggested by the recent data from ATLAS and
CMS at the LHC~\cite{Higgs-LHC}.

We also estimate the axino production from the decay of both Higgsino and saxion.
In our scenario, the dark matter is made of the axion in the form of
non-relativistic coherent field oscillations and the axino lightest SUSY particle
(LSP), while the axion produced by the saxion decay accounts for the dark radiation.
Thus, the axion and its superpartners play an important role to account for
the dark matter and dark radiation.

\section{PQ Extension of the MSSM}

\subsection{Saxion Properties}

Let us begin by examining the properties of the saxion in a simple
PQ extension of the minimal SSM (MSSM) where
the PQ sector consists of the axion
superfield $S$ and $N_\Psi$ pairs of PQ messengers $\Psi+\bar\Psi$
forming ${\bf 5+\bar 5}$ representation of SU(5):
\bea
W_{\rm PQ} = y_\Psi S \Psi\bar\Psi.
\eea
Since the saxion is a flat direction in the supersymmetric limit,
its properties are determined by how SUSY breaking is transmitted
to the PQ sector.
It is natural to expect that the PQ messengers, which are charged
under the SM gauge groups, feel SUSY breaking in the same way as
the MSSM superfields do.

The PQ messengers radiatively generate a potential for the saxion
after SUSY breaking \cite{Coleman:1973jx}, as well as an effective
coupling of $S$ to the gluon supermultiplet that implements the
PQ mechanism.
To see this, one can integrate out $\Psi+\bar\Psi$ under a large
background value of $S$, which results in
\bea
\label{K-S}
{\cal L}_{\rm eff} = \int d^4 \theta Z_S(Q=y_\Psi|S|)|S|^2.
\eea
There then arises a potential, $V_{\rm rad} =m^2_S(Q=y_\Psi|S|)|S|^2$.
Here $m^2_S$ is the soft scalar mass squared of $S$, and $Q$ is
the renormalization scale.
Hence, if $m^2_S$ is positive at a high scale and the messenger
Yukawa coupling is large enough to drive it negative through
radiative correction, $V_{\rm rad}$ develops a minimum along $|S|$
around the scale where $m^2_S$ changes its sign.
At the minimum, the K\"ahler potential (\ref{K-S}) gives masses
to the saxion and axino
\bea
m^2_\sigma &=& \frac{1}{8\pi^2}
\sum_\Psi y^2_\Psi (m^2_\Psi+m^2_{\bar\Psi}+A^2_\Psi)
\sim \frac{5N_\Psi y^2_\Psi}{8\pi^2}m^2_{\rm soft},
\nonumber \\
m_{\tilde a} &=& \frac{1}{8\pi^2}\sum_\Psi y^2_\Psi A_\Psi
\sim \frac{5N_\Psi y^2_\Psi}{8\pi^2}A_\Psi,
\eea
where $m^2_i$ is a soft scalar mass squared, and $A_\Psi$ is the
trilinear soft parameter associated with $y_\Psi$.
The saxion acquires a mass suppressed by a factor
$\epsilon \sim \sqrt{5N_\Psi/8\pi^2}\,y_\Psi$ relative to other
superparticle masses $m_{\rm soft}$, and the axino has a small
mass $m_{\tilde a}\sim \epsilon^2 A_\Psi$.
From the effective K\"ahler potential, one can also find the saxion
couplings
\bea
\frac{\sigma}{\sqrt2 c_aF_a}\left(
(\partial^\mu a)\partial_\mu a +\frac{1}{2}
(\lambda_{\tilde a}m_{\tilde a} \tilde a\tilde a
+ {\rm h.c.}) \right),
\eea
where $F_a=c^{-1}_a\langle |S| \rangle$ is the axion decay constant.
The model-dependent parameter $c_a$ is generally of order unity,
and the saxion coupling to the axino
\bea
\lambda_{\tilde a} = \left.\frac{d\ln(y^2_\Psi A_\Psi)}{d\ln Q}
\right|_{Q=c_aF_a}
\eea
generically has a value in the range between $10^{-1}$ and $10^{-2}$.

On the other hand, depending on the mediation mechanism of SUSY
breaking, $V_{\rm rad}$ alone may not be able to stabilize $S$.
This happens, for instance, in gauge mediation, where only
those carrying SM gauge charges acquire soft SUSY breaking masses
at the messenger scale $M_{\rm mess}$.
An interesting way to give a soft mass to the gauge singlet $S$
at $M_{\rm mess}$ is to consider mixing between $\Psi$ (or $\bar\Psi$)
and the messenger fields transferring SUSY breaking \cite{Jeong:2011xu}.
Then, there is an additional contribution to soft masses for
the PQ sector fields, while the MSSM sector feels SUSY breaking only
through the gauge mediation.
The additional contribution makes $V_{\rm rad}$ develop a minimum
in the same way as discussed above.
Turning off the mixing, the saxion potential runs away to infinity,
but can be lifted by supergravity effects $\Delta V = \xi m^2_{3/2}|S|^2$
where $m_{3/2}$ is the gravitino mass.
For a positive $\xi$ of order unity and $m_{3/2}$ not so small compared
to $m_{\rm soft}$, $S$ is stabilized below $M_{\rm mess}$ by
$V_{\rm rad}+\Delta V$.
The situation is different when $m_{\rm soft}\gg m_{3/2}$, for which
the supergravity effects become important at $|S|$ larger than
$M_{\rm mess}$.\footnote{
In this case, a minimum appears at a scale
above $M_{\rm mess}$, where
$\Psi+\bar\Psi$ generate a potential for the saxion at the three-loop
level \cite{Asaka:1998ns}.
This results in that the saxion has a mass of the order of $m_{3/2}$,
and the axino acquires a tiny mass
$m_{\tilde a}=(\epsilon_1+\epsilon_2)m_{3/2}\ll m_{3/2}$, where
$\epsilon_1 \sim m_{3/2}/(8\pi^2 m_{\rm soft})$ comes from the PQ
messenger loops while $\epsilon_2 \sim F^2_a/M^2_{Pl}$ is due to
the supergravity contribution.
The coupling $\lambda_{\tilde a}$ has a value about
$(\epsilon_1/8\pi^2 + \epsilon_2)/(\epsilon_1+\epsilon_2)$,
and $F_a\sim M_{\rm mess}m_{\rm soft}/(\sqrt{8\pi^2} m_{3/2})$.
}

We close this subsection by stressing that the potential generated by
PQ messenger loops is naturally expected to play an important role in
stabilizing the saxion in models where the PQ symmetry is spontaneously
broken mainly by a single PQ field.
Then, the saxion and axino are generally lighter than the MSSM superparticles.

\subsection{Saxion Cosmology}

Let us move to the cosmology of the saxion.
The saxion has a very flat potential lifted only by the SUSY breaking effect, and so,
the saxion potential may be significantly modified during inflation, because
SUSY is largely broken
by the inflaton potential energy. In particular, the so-called Hubble-induced mass term
generically deviates the saxion from the low-energy potential minimum.\footnote{
This is not the case if there is a certain (approximate) symmetry among the PQ fields.
}

For instance, the saxion may be stabilized at a large field value close to
the Planck scale in the presence of the negative Hubble-induced mass term.
In this case, the saxion starts to oscillate with a large initial amplitude when
the Hubble parameter becomes comparable to the saxion mass $m_\sigma$.
Then the saxion would dominate the energy density of the Universe,
if the reheating temperature of the inflaton is  higher than the saxion decay
temperature.

Alternatively, the PQ symmetry may  be restored during and/or after inflation.
In particular, since the PQ messengers generate a thermal potential for the saxion,
$V_{\rm thermal}\sim y^2_\Psi T^2|S|^2$ at $|S|\ll T$,
the saxion could be thermally trapped at the origin if it sits around the origin
after inflation.
In this case thermal inflation \cite{Lyth:1995hj} takes place when the saxion
potential energy at the origin ($\sim m^2_\sigma F^2_a$) dominates the Universe.

Thus, it is plausible that the saxion dominates the energy density of the Universe.
In the rest of the paper, we assume that this is indeed the case, and discuss
how the saxion decay proceeds.

\subsection{PQ Solution to the $\mu$ Problem}

If $S$ has no other interactions with the MSSM sector than the loop-induced
coupling to the gluon supermultiplet, the saxion dominantly decays into axions
with a branching ratio $B_a\simeq 1$:
\bea
B_a = \frac{\Gamma_{\sigma\to aa}}{\Gamma_\sigma}
= \frac{1}{\Gamma_\sigma} \frac{1}{64\pi}\frac{m^3_\sigma}{(c_aF_a)^2},
\eea
where $\Gamma_\sigma$ is the total decay width of the saxion.
This is obviously problematic in a scenario where the Universe experiences
a saxion-dominated epoch.

A natural way to suppress $B_a$ is to introduce a coupling of $S$ to
the Higgs doublets so that a Higgs $\mu$ term is dynamically generated
after PQ symmetry breaking.
Among various types of couplings, we take the Kim-Nilles superpotential
term \cite{Kim:1983dt}
\bea
\label{KN}
W = \lambda\frac{S^2}{M_{Pl}}H_uH_d,
\eea
which generates an effective $\mu$ term
\bea
\mu = \lambda \frac{(c_aF_a)^2 }{M_{Pl}},
\eea
around the weak scale for $\lambda\lesssim 1$ and $F_a=10^{10-12}$ GeV,
and thus naturally explains the smallness of $\mu$.\footnote{
To solve the $\mu$ problem, one may instead consider the Giudice-Masiero
mechanism \cite{Giudice:1988yz} implemented by a K\"ahler potential term
$S^*H_uH_d/S$.
In this case, however, the couplings of the saxion to the MSSM
sector arise at the loop level.
}
The above superpotential term also induces couplings of the saxion
to the MSSM particles.
These couplings are proportional to $\mu/F_a$, and open up the
possibility to obtain $\Delta N_{\rm eff}\sim 1$ from the axions
non-thermally produced by saxion decays.

To evaluate $B_a$, one needs to know interactions of the saxion with
the MSSM particles.
For simplicity, we consider the decoupling limit where the effective Higgs
sector below $m_{\rm soft}$ consists only of a SM-like Higgs doublet and
other heavy Higgs bosons decouple from the theory.
Then, the saxion has the interactions
\bea
\frac{C_\sigma}{\sqrt2}
\frac{|\mu|^2}{c_aF_a}\left(1-\frac{|B|^2}{m^2_A}\right)
\sigma h h
+
\left(
\frac{C_\sigma}{\sqrt2}
\frac{\mu}{c_aF_a} \sigma \tilde H_u \tilde H_d
+ {\rm h.c.}
\right),
\eea
both of which are directly induced from (\ref{KN}).
Here $h$ is the CP-even neutral Higgs boson, $m_A$ is the mass of
the CP-odd neutral Higgs boson, and $B$ is the Higgs mixing soft
parameter.
The coefficient $C_\sigma=\partial \ln|\mu|/\partial \ln|S|=2$
reflects the form of the coupling between $S$ and the Higgs doublets.
In addition, the saxion interacts with other MSSM particles as well
because there arises mixing between the saxion (axino) and neutral
Higgs (Higgsinos) after electroweak symmetry breaking.
These couplings can be read off from the MSSM Lagrangian by taking
the substitution
\bea
h &\to& \frac{2C_\sigma v}{c_aF_a}\left(1-\frac{|B|^2}{m^2_A}\right)
\frac{|\mu|^2}{m^2_\sigma-m^2_h} \sigma,
\eea
where $\langle |H^0_u|\rangle = v\sin\beta$.
It is straightforward to derive the partial decay widths of the saxion
into SM particles and Higgsinos, which are presented in the appendix.
The important decay channels are those into $hh$, $WW$, $ZZ$, $b\bar b$,
$t\bar t$, and into a Higgsino pair, depending on $m_\sigma$.
We note that $B_a$ is determined essentially by $\mu$ and $m_\sigma$,
but insensitive to $F_a$.
The saxion interacts with the SM particles more strongly for a larger $\mu$,
making $B_a$ smaller.

\section{Axion Dark Radiation}

In this section, we examine the condition for the axion dark radiation
produced at saxion decays to yield $\Delta N_{\rm eff}\sim 1$
in the presence of the superpotential term (\ref{KN}) that is
responsible for generating a $\mu$ term.
Then, we also examine if the axino can constitute the dark matter of the
Universe with the correct amount of a relic density.

Our analysis does not assume any specific mechanism of the saxion
stabilization, but the radiative potential induced by the PQ messengers
may play an important role in fixing the vacuum expectation value of $S$.
Keeping this in mind, we shall consider the case where the saxion and axino
have masses lighter than $m_{\rm soft}$.
It should be noted that the Higgsino mixing parameter $\mu$, on which $B_a$
crucially depends, does not need to be related to the scale $m_{\rm soft}$
because it is a supersymmetric coupling.

\subsection{Axion Dark Radiation from Saxion Decays}

The axions produced at the saxion decays have never been in thermal
equilibrium for the saxion decay temperature much lower than $F^2_a/M_{Pl}$
\cite{Choi:1996vz}, which is the case we shall deal with.
The energy density of non-thermalized axions and thermalized radiation evolves
as
\bea
\rho_a(t) &\propto& a^{-4}(t),
\nonumber \\
\rho_{\rm SM}(t) &\propto& g^{-1/3}_\ast(t)a^{-4}(t),
\eea
where $a(t)$ is the scale factor, and $\rho_a:\rho_{\rm SM}=B_a:(1-B_a)$ at
the saxion decay time $t=t_\sigma$.
Because non-thermally produced axions contribute to $N_{\rm eff}$ with
\bea
\Delta N_{\rm eff} &=&  \frac{\rho_a}{\rho_\nu}\,
\Big|_{\,\nu\,{\rm decouple}}
= \frac{\rho_{\rm SM}(t)}{\rho_\nu(t)}
\,\Big|_{\,\nu\,{\rm decouple}}\times
\frac{\rho_a(t)}{\rho_{\rm SM}(t)}\,
\Big|_{\,\nu\,{\rm decouple}}
\nonumber \\
&=& \frac{43}{7} \frac{B_a}{1-B_a}
\left(\frac{43/4}{g_\ast(T_\sigma)}\right)^{1/3},
\eea
$\Delta N_{\rm eff}=1$ is achieved if the saxion couplings to the MSSM sector
arising from the superpotential (\ref{KN}) are strong enough to give $B_a$
between about $0.2$ and $0.3$.
Here $\rho_\nu$ is the energy density of a single species of relativistic
neutrino, and
the effective number of relativistic degrees of freedom varies
from $g_\ast\simeq 60$ to $g_\ast=113.75$ for $0.2\,{\rm GeV}<T<m_{\rm soft}$,
through which a mild dependence on $T_\sigma$ comes in.
The saxion decay temperature $T_\sigma$ is defined at the time
$t=t_\sigma$ when the energy density of radiation $\rho_a+\rho_{\rm SM}$ becomes
equal to that of the saxion, which implies
$\rho_{\rm SM}(t_\sigma)=(1-B_a) \rho_\sigma(t_\sigma)$,
where we have used that the ratio between energy densities of axion and
thermalized radiation is $\rho_a/\rho_{\rm SM} = B_a/(1-B_a)$ at $t=t_\sigma$.
One can then estimate,
\bea
T_\sigma &\simeq& \left(\frac{45(1-B_a)}{4\pi^2 g_\ast(T_\sigma)}
\right)^{1/4}\sqrt{\Gamma_\sigma M_{Pl}}
\nonumber \\
&\simeq&
3.5{\rm GeV}\left(
\frac{B^2_a/(1-B_a)}{0.08} %\frac{1-B_a}{B^2_a}
\right)^{-1/4}
\left(\frac{g_\ast(T_\sigma)}{100}\right)^{-1/4}
\left(\frac{m_\sigma}{300{\rm GeV}}\right)^{3/2}
\left(\frac{c_a F_a}{10^{11}{\rm GeV}}\right)^{-1},
\eea
with $t_\sigma\simeq 1/\Gamma_\sigma$, by taking the approximation
that the scale factor $a(t)$ of the Universe is determined mainly by
the saxion energy density at $t<t_\sigma$ and by the energy density
of the radiation at $t>t_\sigma$.

\begin{figure}[t]
\begin{center}
\begin{minipage}{16.4cm}
\centerline{
{\hspace*{0cm}\epsfig{figure=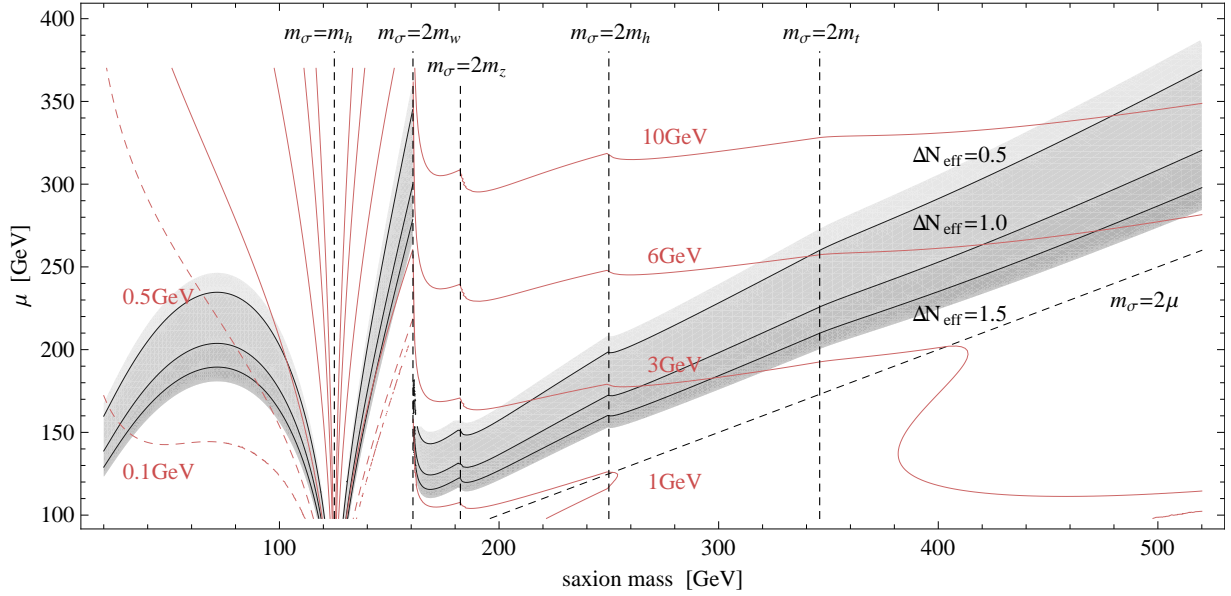,angle=0,width=16.2cm}}
%{\hspace*{.4cm}\epsfig{figure=.eps,angle=0,width=7.2cm}}
}
\caption{The constant contours of $\Delta N_{\rm eff}$
for $|B|/m_A=0.6$ and $m_h=125$ GeV in the $(m_\sigma,\mu)$ plane.
The black lines represent the contours of
$\Delta N_{\rm eff}=0.5,\,1,\,1.5$ from the above, respectively.
In the shadowed region, $0.4\leq \Delta N_{\rm eff}\leq 2$.
We also plot constant contours of the quantity,
$(c_aF_a/10^{11}{\rm GeV})\times T_\sigma=0.1,\,0.5,\,1,\,3,\,6,\,10$
GeV from the below, respectively, in red lines.
}
\label{fig:DR}
\end{minipage}
\end{center}
\end{figure}

In fig. \ref{fig:DR}, we show constant contours of $\Delta N_{\rm eff}$
in the $(m_\sigma,\mu)$ plane for $|B|/m_A=0.6$ and $m_h=125$ GeV.
Here we take $g_\ast=80$, which depends on $T_\sigma$ but
modifies $\Delta N_{\rm eff}$ only slightly for $0.2{\rm GeV}<T_\sigma<m_{\rm soft}$.
It is interesting to observe that $\Delta N_{\rm eff}\sim 1$ is
obtained at $\mu$ of a few hundred GeV for $m_\sigma$ less than 1 TeV.
A large $\mu$ renders the saxion couplings to the SM particles strong,
suppressing $B_a$.
We find that non-thermally produced axions yield $\Delta N_{\rm eff}=1$ at
$\mu$ similar to the saxion mass:
\bea
\mu \sim
150{\rm GeV}\times \left|\,1-\frac{|B|^2}{m^2_A}\right|^{-1/2}
\left(\frac{m_\sigma}{300{\rm GeV}}\right),
\eea
for $2m_W < m_\sigma \lesssim m_{\rm soft}$, in which case the saxion decays
mainly through $\sigma\to WW$ and through $\sigma \to ZZ,\,hh,\,t\bar t$ if
kinematically allowed.
Here $m_W$ is the $W$ boson mass.
On the other hand, for $m_\sigma<2m_W$, where the dominant decay channel
is $\sigma\to b\bar b$, a $\mu$ less than about 300 GeV is necessary to
have $\Delta N_{\rm eff}=1$.
In the figure, we also draw constant contours of the quantity
\bea
\left(\frac{c_aF_a}{10^{11}{\rm GeV}}\right)T_\sigma,
\eea
from which the value of $T_\sigma$ can be read off for a given value
of $F_a$.
As we will see soon, the dark matter abundance puts a stringent
constraint on $T_\sigma$ and also on the properties of the axino.

The saxion generically acquires a mass smaller than $m_{\rm soft}$
by one order of magnitude from the potential radiatively generated
by the PQ messengers.
Meanwhile, the recent data from the LHC Higgs search suggests
that a SM-like Higgs boson may have mass around 125 GeV
\cite{Higgs-LHC}.
To explain this within the MSSM, we need to push the stop mass to
about 10 TeV or higher, or to invoke large stop mixing.
For $m_\sigma\lesssim 0.1m_{\rm soft}$ and
$1{\rm TeV}\lesssim m_{\rm soft} \lesssim 10{\rm TeV}$,
the Higgsino should have a mass of a few hundred GeV when the axion
dark radiation gives $\Delta N_{\rm eff}\sim 1$.
Thus, there is a chance to detect SUSY at multi-TeV hadron
colliders even when the other MSSM superparticles are heavier than 1 TeV.

\subsection{Axino Dark Matter}

Under the $R$-parity conservation, the axino is a natural candidate
for the dark matter in PQ extensions of the MSSM.
The main processes of axino production for $\mu<m_{\rm soft}$
are $(i)$ the decay of Higgsinos in thermal bath,
and $(ii)$ the decay of the saxion, which crucially depends on
the properties of the axino.

Let us first examine the thermal process $(i)$, which is mediated
by the interactions
\bea
\frac{C_{\tilde a}}{\sqrt 2}\frac{\mu}{c_aF_a} h \left(\tilde H^0_u \cos\beta
+ \tilde H^0_d \sin\beta\right) \tilde a
+ \frac{C_{\tilde a}}{\sqrt 2}\frac{m_Z}{c_aF_a}
Z_\mu\left( \bar{\tilde H}^0_u\sin\beta
- \bar{\tilde H}^0_d\cos\beta \right) \sigma^\mu\tilde a,
\eea
with $C_{\tilde a}= \partial\ln \mu/\partial \ln S =2$,
where the first coupling comes from the superpotential term $S^2H_uH_d$,
while the other is a consequence of the axino-Higgsino mixing.
The produced axino number density is highly sensitive to $T_\sigma$,\footnote{
As was noticed in Ref. \cite{Chun:2011zd}, in the case that the Universe does not
experience the saxion domination, thermally produced axinos by Higgsino
decays would overclose the Universe unless
the reheating temperature is much lower than the weak scale or
the axino has a tiny mass less than ${\cal O}(100)$ keV for
$F_a\sim 10^{11}$ GeV.
See also Refs. \cite{Bae:2011jb,Choi:2011yf,Bae:2011iw} for thermal production
of the axino.
}
and is numerically approximated by $n_{\tilde a}/s \propto e^{-0.63 \mu/T_\sigma}$
for $T_f\lesssim T_\sigma\lesssim 0.2\mu$ where $T_f\simeq \mu/20$ is
the freeze-out temperature of the Higgsino.
The derivation can be found in the appendix.
We find that $n_{\tilde a}/s\leq 10^{-10}$ requires $T_\sigma\lesssim T_f$
for $\mu\sim m_\sigma$ and $B_a\sim 0.1$.
For $T_\sigma\leq T_f$, the axino energy density is approximated by
\bea
\label{thermal-axino}
\left.\frac{\rho_{\tilde a}}{s}\right|_{\rm thermal}
&=& \frac{m_{\tilde a}}{s} \sum^2_{i=1} \left( \frac{1}{a^3}
\int^{t_f}_0 dt\, a^3(t) \Gamma_{\tilde H^0_i} n_{\tilde H^0_i}(t)
+ n_{\tilde H^0_i}(t_f) \right)
\nonumber \\
&\approx& 3.6\times 10^{-10}{\rm GeV}
\left(\frac{m_{\tilde a}}{1{\rm GeV}}\right)
\left(\frac{g_\ast(T_\sigma)}{100}\right)^{-1}
\left(\frac{B_a}{0.3}\right)
\left(\frac{\mu/m_\sigma}{0.5}\right)^3
\left(\frac{T_\sigma/\mu}{1/34}\right)^9,
\eea
where $\Gamma_{\tilde H^0_i}$ is the decay rate for $\tilde H^0_i \to h(Z)+\tilde a$,
$n_{\tilde H^0_i}$ is the number density, and $t_f$ is the time when
the Higgsino freezes out of thermal equilibrium.
We have taken into account that there are two neutral Higgsinos
$\tilde H^0_{1,2}$, which are almost degenerate in mass, and neglected mixing
between Higgsinos and gauginos.
Meanwhile, because $m_\sigma< 2\mu$ is required for
$\Delta N_{\rm eff}\leq 2$ unless $B$ is much smaller than $m_{\rm soft}$,
the saxion decay into a Higgsino pair is kinematically forbidden.
Even if this mode is open, the annihilation among the Higgsinos produced
at the saxion decays would occur effectively \cite{Hisano:2000dz}, and consequently
the axino relic abundance produced from the Higgsino decays can be smaller
than the observed dark matter abundance for $m_{\tilde a}$ less than $T_\sigma$.

On the other hand, the non-thermal process $(ii)$ crucially depends on
the properties of the axino.
This process gives rise to
\bea
\label{non-th-axino}
\left.\frac{\rho_{\tilde a}}{s}\right|_{\rm non-th}
&=& \frac{2\Gamma_{\sigma \to\tilde a\tilde a}}{a^3 s}
\frac{m_{\tilde a}}{m_\sigma}
\int^\infty_0 dt\, a^3(t) \rho_\sigma(t)
\nonumber \\
&\approx&
\frac{3e^{\Gamma_\sigma t_\sigma}}{2}
\frac{T_\sigma}{1-B_a}
\frac{m_{\tilde a}}{m_\sigma}
\frac{\Gamma_{\sigma \to\tilde a\tilde a}}{\Gamma_\sigma}
\nonumber \\
&\simeq&
2.4\times 10^{-10}{\rm GeV}
\left(\frac{\lambda_{\tilde a}}{0.01}\right)^2
\left(\frac{B_a/(1-B_a)}{0.3}\right)
\left(\frac{m_{\tilde a}/m_\sigma}{0.01}\right)^3
\left(\frac{T_\sigma}{1{\rm GeV}}\right).
\eea
Thus, in order not to overclose the Universe, the axino should have
a small coupling to the saxion and/or small mass compared to $m_\sigma$.
The non-thermally produced axinos can yield $\Omega_{\tilde a}h^2\simeq 0.1$,
for instance, if $m_{\tilde a}$ is less than a few GeV and
$\lambda_{\tilde a}\lesssim 10^{-2}$ for $F_a=10^{10-12}$ GeV.
Such axino properties are indeed expected when $S$ is stabilized by
the potential generated from the PQ messenger loops.

The axino abundance, (\ref{thermal-axino}) and (\ref{non-th-axino}), can
be made consistent with the observed value of the dark matter abundance
by taking appropriate values of $m_{\tilde a}$, $\lambda_{\tilde a}$
and $F_a$, for $\mu$ and $m_\sigma$ leading the axion dark radiation
to give $\Delta N_{\rm eff}\sim 1$.
The energy density of the dark matter receives contribution also from
the axion due to the vacuum misalignment
\bea
\Omega_a h^2 \sim 0.4 \theta^2_a
\left(\frac{F_a}{10^{12}{\rm GeV}}\right)^{1.18},
\eea
where $|\theta_a|<\pi$ is the initial misalignment angle.
If the axino relic abundance is too small, which would be the case
for $F_a$ around $10^{12}$ GeV or higher, the dark matter of the Universe
can be explained by the axion from the misalignment.

One might consider other cases where the axino is heavier than the Higgsino
or some other MSSM superparticle.
Then, for $\mu$ of a few hundred GeV as required to have
$\Delta N_{\rm eff}\sim 1$, one needs either $m_\sigma <2m_{\tilde a}$ or
an extremely small $\lambda_{\tilde a}$ in order to avoid overproduction
of the dark matter.
In the case where the gravitino is the lightest superparticle, which is possible
in gauge mediation, a small gravitino mass $m_{3/2} \ll m_{\tilde a}$
or a tiny $\lambda_{\tilde a}$ would be necessary since gravitinos produced
at the axino decays behave like a hot dark matter with a free-streaming length
much larger than 10 Mpc.

Finally, we mention the detection potential of SUSY at collider experiments.
The charged Higgsino generally obtains a mass slightly heavier than the mass
of $\tilde H^0_1$ when bino and wino masses have the same phase
\cite{Haber:1984rc,Pierce:1996zz}.
Assuming that it is the lightest one of the MSSM superparticles,
$\tilde H^0_1$ decays into $h(Z)+\tilde a$ with
\bea
\Gamma_{\tilde H^0_1} \approx
\frac{C^2_{\tilde a}}{16\pi}\frac{\mu^3}{(c_aF_a)^2}
\simeq \frac{1}{314{\rm cm}}\left(\frac{\mu}{200{\rm GeV}}\right)^3
\left(\frac{c_aF_a}{10^{11}{\rm GeV}}\right)^{-2},
\eea
for $\mu>m_h+m_{\tilde a}$, as is the case in most of the parameter region
giving $\Delta N_{\rm eff}\leq 2$ for $m_h\lesssim 130$ GeV.
Here we have neglected the masses of the final states and mixing with
neutral gauginos.
Thus, depending on $\mu$ and $F_a$, $\tilde H^0_1$ can decay inside the detector
while leaving displaced vertices.
Measuring its decay length would give us information about the axion
decay constant.

\section{Conclusions}
In this paper we have studied the possibility that dark radiation recently suggested
by observations can be explained by the QCD axion non-thermally produced by the
saxion decay. In order to account for $\Delta N_{\rm eff}\sim 1$, the axion superfield
must have a sizable coupling to the Higgs sector.
We have found that the Higgsino mixing parameter $\mu$ is bounded
above and should be in the range of a few hundred GeV, for the saxion mass lighter
than $1$\,TeV. Considering that the saxion mass could be naturally one order of magnitude
smaller than the soft masses for the MSSM superparticles, the Higgsino can be within
the reach of the LHC  and/or ILC, even if the other SUSY particles are much heavier
than $O(1)$\,TeV. This will be of great importance especially if the SM-like Higgs boson
mass is confirmed to be around $124-126$\,GeV~\cite{Higgs-LHC}.

%%%%%%%%%%%%%%%%%%%%%%%%%%%%%%%%%%%%%
\section*{Acknowledgment}
%%%%%%%%%%%%%%%%%%%%%%%%%%%%%%%%%%%%%
This work was supported by the Grant-in-Aid for Scientific Research on Innovative
Areas (No. 21111006) [FT], Scientific Research (A)
(No. 22244030 and No.21244033 [FT]), and JSPS Grant-in-Aid
for Young Scientists (B) (No. 21740160) [FT].  This work was also
supported by World Premier International Center Initiative (WPI
Program), MEXT, Japan, and by Grants-in-Aid for Scientific Research from the Ministry
of Education, Science, Sports, and Culture (MEXT), Japan (No. 23104008 and No.
23540283 [KSJ]).

\section*{Saxion decay rates and axino production}

In this appendix, we present the partial decay widths of the saxion,
and the derivation of the axino abundance thermally produced by Higgsino
decays.
In the decoupling limit $m_A\gg m_W$, the saxion decay occurs with
\bea
\Gamma_{\sigma\to hh} &=&
\left(1-4\frac{m^2_h}{m^2_\sigma}\right)^{1/2} \Lambda_\sigma,
\nonumber \\
\Gamma_{\sigma\to VV} &=&
k_V  \frac{m^4_\sigma}{(m^2_\sigma-m^2_h)^2}
\left(1-4\frac{m^2_V}{m^2_\sigma}\right)^{1/2}
\left(1-4\frac{m^2_V}{m^2_\sigma}
+12\frac{m^4_V}{m^4_\sigma}\right) \Lambda_\sigma,
\nonumber \\
\Gamma_{\sigma\to f\bar f} &=&
4N_f \frac{ m^2_f m^2_\sigma}{(m^2_\sigma-m^2_h)^2}
\left(1-4\frac{m^2_f}{m^2_\sigma} \right)^{3/2} \Lambda_\sigma,
\eea
if the corresponding process is kinematically accessible.
Here $k_V=2$(1) for $V=W(Z)$, and $N_f=3(1)$ for quarks (leptons).
The overall factor $\Lambda_\sigma$ is defined by
\bea
\Lambda_\sigma = 4C^2_\sigma B_a \Gamma_\sigma \left(
1-\frac{|B|^2}{m^2_A}\right)^2\frac{|\mu|^4}{m^4_\sigma},
\eea
where we have used the relation $m^2_A=2|B\mu|/\sin2\beta$.
On the other hand, if $m_\sigma>2\mu$, the saxion decays
into a pair of Higgsinos with
\bea
\Gamma_{\sigma\to \tilde H \tilde H}
= 8C^2_\sigma B_a\Gamma_\sigma \frac{|\mu|^2}{m^2_\sigma}
\left(1-4\frac{|\mu|^2}{m^2_\sigma} \right)^{3/2},
\eea
ignoring mixing between the Higgsinos and gauginos.
The above decay rate will be reduced by the mixing.

Let us move to the axino abundance.
The energy densities of the saxion and thermalized radiation are
given by
\bea
\rho_\sigma(t) &=& \rho_0 \left(\frac{a_0}{a(t)}\right)^3
e^{-\Gamma_\sigma(t-t_\sigma)},
\nonumber \\
\rho_{\rm SM}(t) &=&
(1-B_a)\Gamma_\sigma \rho_0 \left(\frac{a_0}{a(t)}\right)^4
\int^t_{t_\sigma}
dt^\prime \frac{a(t^\prime)}{a_0}e^{-\Gamma_\sigma(t^\prime-t_\sigma)},
\eea
where $a_0=a(t_\sigma)$, $\rho_0=\rho_\sigma(t_\sigma)$.
Thermal production of axinos is dominated by the Higgsino decays:
\bea
\Gamma_{\tilde H^0_{1,2}\to h\tilde a} &\simeq&
(\cos\beta\pm\sin\beta)^2
\frac{C^2_{\tilde a}}{32\pi}
\frac{|\mu|^3}{(c_aF_a)^2}\left(1-\frac{m^2_h}{|\mu|^2}\right)^2,
\nonumber \\
\Gamma_{\tilde H^0_{1,2}\to Z\tilde a} &\simeq&
(\cos\beta\mp\sin\beta)^2
\frac{C^2_{\tilde a}}{32\pi}
\frac{|\mu|^3}{(c_aF_a)^2}\left(1-\frac{m^2_Z}{|\mu|^2}\right)^2
\left(1+2\frac{m^2_Z}{|\mu|^2}\right),
\eea
for $\mu>m_{h,Z}$ and $\mu\gg m_{\tilde a}$.
The number density of thermally produced axinos is thus estimated by
\bea
\frac{n_{\tilde a}}{s}
&=& \sum_i \left(
\frac{\Gamma_{\tilde H^0_i\to h\tilde a}+\Gamma_{\tilde H^0_i\to Z\tilde a}}{a^3s}
\int^{t_f}_0 dt\, a^3(t) n_{\tilde H^0_i}(t)
+ \frac{n_{\tilde H^0_i}(t_f)}{s} \right)
\nonumber \\
&\simeq&
\frac{45}{\pi^4 g_\ast(T_\sigma)} \left(
\sum_i \frac{\Gamma_{\tilde H^0_i\to h\tilde a, Z\tilde z}}
{\Gamma_\sigma}
( P_{t<t_\sigma}\left(z_\sigma)+ P_{t>t_\sigma}(z_\sigma)\right)
\right.
\nonumber \\
&&
\left.
\hspace{2cm}
+\, z^{-3}_f \int^\infty_0 dk \frac{k^2}{e^{\sqrt{k^2+1}/z_f}+1}
\times {\rm Min}[ (z_\sigma/z_f)^5,1 ]
\right),
\eea
where $z_f=T_f/\mu$ and $z_\sigma=T_\sigma/\mu$.
The axino production functions $P_i(x)$ before and after $t=t_\sigma$
are approximated by
\bea
P_{t<t_\sigma}(x) &\simeq&
2 x^9 \int^\infty_{x_0}dz
\int^\infty_0dk \frac{z^{-13}k^2}{e^{\sqrt{k^2+1}/z}+1}
\approx \left(\frac{x}{x_0}\right)^9
e^{-0.63/x_0},
\quad
\nonumber \\
P_{t>t_\sigma}(x) &\simeq& \theta(x-z_f)\, x^2
\int^x_{z_f} dz \int^\infty_0 dk \frac{z^{-6}k^2}{e^{\sqrt{k^2+1}/z}+1}
\approx \theta(x-z_f)\, x^{-2}e^{-1.26/x},
\eea
for $x_0={\rm Max}[x,z_f]$, and $\theta(x)$ being the step function.
Here we have taken the approximation,
$a(t)/a_0\sim (t/t_\sigma)^{2/3} \sim (T_\sigma/T)^{8/3}$ at $t<t_\sigma$.
For $z_\sigma < 0.2$, $P_{t<t_\sigma}(z_\sigma)$ is always
larger than $P_{t>t_\sigma}(z_\sigma)$.
We also find that the axino abundance computed by the approximated
formulae is slightly smaller by a factor 2 or 3
than the one obtained by numerically solving the Boltzmann equation.
However, the above approximation is enough to illustrate how strongly
the axino abundance depends on $T_\sigma$.

\end{document}